# Boosting radiotherapy dose calculation accuracy with deep learning

Yixun Xing, Ph.D.*, You Zhang, Ph.D.*, Dan Nguyen, Ph.D., Mu-Han Lin, Ph.D., Weiguo Lu, Ph.D., Steve Jiang, Ph.D.
Medical Artificial Intelligence and Automation (MAIA) Laboratory, Department of Radiation Oncology, University of Texas Southwestern Medical Center, Dallas, TX 75390, USA
* Yixun Xing and You Zhang contributed equally to the work.

**Corresponding address:**
You Zhang, Ph.D.
Department of Radiation Oncology
UT Southwestern Medical Center
2280 Inwood Road, Dallas, TX 75390
Phone: 214-645-2699
Email: you.zhang@utsouthwestern.edu

Steve Jiang, Ph.D.
Department of Radiation Oncology
UT Southwestern Medical Center
2280 Inwood Road, Dallas, TX 75390
Phone: 214-645-8510
Email: steve.jiang@utsouthwestern.edu

**Running title:**
Boosting dose accuracy with deep learning

# Abstract

In radiotherapy, a trade-off exists between computational workload/speed and dose calculation accuracy. Calculation methods like pencil-beam convolution can be much faster than Monte-Carlo methods, but less accurate. The dose difference, mostly caused by inhomogeneities and electronic disequilibrium, is highly correlated with the dose distribution and the underlying anatomical tissue density. We hypothesize that a conversion scheme can be established to boost low-accuracy doses to high-accuracy, using intensity information obtained from computed tomography (CT) images. A deep learning-driven framework was developed to test the hypothesis by converting between two commercially-available dose calculation methods: AAA (anisotropic-analytic-algorithm) and AXB (Acuros XB). A hierarchically-dense U-Net model was developed to boost the accuracy of AAA dose towards the AXB level. The network contained multiple layers of varying feature sizes to learn their dose differences, in relationship to CT, both locally and globally. AAA and AXB doses were calculated in pairs for 120 lung radiotherapy plans covering various treatment techniques, beam energies, tumor locations, and dose levels. For each case, the CT and the AAA dose were used as the input and the AXB dose as the 'ground-truth' output, to train and test the model. The mean-squared-errors (MSEs) and gamma-passing-rates (2mm/2% & 1mm/1%) were calculated between the boosted AAA doses and the 'ground-truth' AXB doses. The boosted AAA doses demonstrated substantially improved match to the 'ground-truth' AXB doses, with average($\pm$s.d.) gamma-passing-rate (1mm/1%) 97.6%($\pm$2.4%), compared to 87.8%($\pm$9.0%) of the original AAA doses. The corresponding average MSE was 0.11($\pm$0.05) vs 0.31($\pm$0.21). Deep learning is able to capture the differences between dose calculation algorithms to boost the low-accuracy algorithms. By combining a less accurate dose calculation algorithm with a trained deep learning model, dose calculation can potentially achieve both high accuracy and efficiency.

**Key words:** Dose calculation, Deep learning, CT, Inhomogeneous regions, AAA, AXB

# I.   Introduction

In radiation therapy, the radiation doses to the tumor and surrounding normal tissues directly determine treatment efficacy and safety [1, 2]. It is pivotal for the radiation therapy treatment planning systems (TPSs) to accurately calculate the dose distributions to aid physician's decisions. Accurate dose calculation is also key to a reliable and reproducible model between dose distributions and clinical outcomes to guide future treatments [3]. The dose calculation algorithms have seen generations of development. The early generations of algorithms, usually referred to as *correction-based* methods [4-7], are barely physics principle-driven. Their accuracy is highly unreliable in heterogeneous regions (for instance, areas with lung tissue-surrounding tumors), where the loss of electronic equilibrium occurs [8]. To better model the physics in these regions, *model-based* techniques were developed [9-12]. These techniques model the radiation energy transport via dose kernels and convolutions. To account for the tissue heterogeneity, the dose kernels are scaled based on the equivalent electron density path lengths encountered by radiation beams, leading to the superposition-convolution type of algorithm, which is widely-used in today's clinic. Analytical anisotropic algorithm (AAA), one of such algorithms, is commercially implemented in the Eclipse TPS [12] (Varian Medical Systems, Palo Alto, CA). However, a discrepancy over 5% from measurements can still be observed for AAA in inhomogeneous regions [13], which can be clinically significant [14, 15]. The discrepancy is due to the fact that kernel scaling does not explicitly and realistically model the energy transport through physical interactions either. The Monte Carlo algorithm represents a third type of dose calculation algorithms. It models the transport and energy deposition of each particle (photons, electrons, etc.) via explicit physics principles, which are modeled with measured data or proven formula, and provides the highest accuracy [16, 17]. However, Monte Carlo needs to simulate the transport of each particle individually, which requires substantial computational power and may significantly prolong the dose calculation time. Besides Monte Carlo, recently a new dose calculation technique using the linear Boltzmann transport equation was implemented as the Acuros XB (AXB) algorithm in Eclipse [18, 19]. It is proven that AXB would theoretically converge to the same solution as the Monte Carlo algorithm [20]. The accuracy of AXB has been extensively validated [17, 21, 22]. The efficiency of AXB is plan-dependent, which improves with increasing beam numbers (relatively) and favors volumetric modulated arc therapy plans. In some scenarios, however, AXB can be 10 times slower than AAA [23].

In general, a trade-off exists on dose calculation: more accurate dose calculation requires more computational power, and is generally more time consuming and resource demanding. Due to this trade-off, current radiotherapy TPSs may have to use less accurate methods for dose calculation, in order to improve efficiency, especially during plan optimization [24]. Such an adoption is less ideal, since these low-accuracy dose calculations used during optimization may potentially trap the optimization into a local optimum, and yield a sub-optimal final plan. As the radiotherapy society is pursuing more precise, individual-tailored treatments, the use of low-accuracy dose calculation algorithms for plan optimization may fail to generate high-quality plans within a tight time frame, especially for on-line adaptive radiotherapy [25]. A technique enabling dose calculation with both high accuracy and efficiency is thus much desired for plan optimization. In addition, recent advancements of real-time imaging techniques also call for such a technique [26], which will make possible on-the-fly dose monitoring and intervention through real-time plan re-optimization and adaptation.

As mentioned, the differences between low-accuracy and high-accuracy doses are mostly within inhomogeneous regions. The inhomogeneity, however, is fully-captured in simulation CT images that are used for dose calculation. From this observation, we hypothesize that the differences between dose algorithms can be learned and correlated with the dose distribution and the CT intensity information. With this learnt correlation, we can then quickly boost the low-accuracy doses to high-accuracy, to overcome the trade-off between dose calculation accuracy and efficiency. Recently, the developments and applications of artificial intelligence (AI) in radiation therapy have seen tremendous growth [27-32].

Sophisticated convolutional neural networks can handle intensive tasks including medical image de-noising, segmentation, treatment plan optimization/evaluation, and clinical outcome prediction. Some networks, including the U-Net [33], can perform voxel-wise prediction and mapping, which allows a potential voxel-to-voxel dose map conversion to boost the accuracy of low-accuracy doses. Additionally, the U-Net can extract both global and local features from dose distributions and CT images, which can be directly correlated with dose differences between algorithms, as the energy transport is essentially determined by both long-range (global) photon transport and short range (local) electron transport. With dedicated graphics processing units (GPUs), the inference of U-Net can also be executed within seconds, meeting the efficiency requirement. Driven by our hypothesis, in this study we introduced an AI-based framework to achieve rapid, direct 3D dose map conversion from low-accuracy doses to high-accuracy doses. We trained and evaluated the whole framework on AAA ("low-accuracy") and AXB ("high-accuracy") dose maps to demonstrate its effectiveness, since these two algorithms are well-studied, widely-available and can be easily evaluated by other groups. Note that the "low-accuracy" and "high-accuracy" here were defined relatively, since under different context and scenarios, AAA can also be high-accuracy (for instance when compared with pencil-beam convolution), and AXB may be low-accuracy when compared with a full-fledged Monte-Carlo package. The "high" and "low" here thus were determined relatively between the two algorithms under study. We derived and tested the dose conversion model using a large lung cancer patient database, aiming to improve the dose calculation of lung cancer treatment, whose accuracy is the most susceptible to tissue inhomogeneity [28].

## II.　　Materials and Methods

**II.1. Data preparation**

In this study, we retrospectively collected a total of 120 lung cancer patient cases in our institution treated between 06/2017 and 03/2018. The retrospective study was approved under an institutional review board umbrella protocol. All patients were planned in Eclipse V15.5, by techniques ranging from 3D non-coplanar conformal static beams, intensity modulated static beams, 3D conformal arcs to volumetric modulated arcs. The total prescription doses ranged from 24 Gy to 60 Gy, covering both conventional and stereotactic body radiation therapy treatments. The tumors were distributed across both central and peripheral lung regions. Both primary lung tumors and metastatic tumors from breast, liver, kidney and prostate were included. The treatment plans used beam energies ranging from 6 MV, 10 MV, 6 MV FFF to 10 MV FFF. All treatments were designed and successfully delivered on an Elekta VersaHD LINAC with a 160-leaf Agility multi-leaf-collimator head [34]. All cases were planned and treated using AAA as the dose calculation engine, with heterogeneity correction turned on. The dose grid was 2.5 mm x 2.5 mm x 2.0 mm in resolution.

For each AAA dose distribution, we calculated the corresponding AXB dose distribution under the exact same plan. The AXB doses were reported in the form of dose to medium to account for the elemental composition of different tissues. In Eclipse, the tissue designation is based on the densities determined from CT Hounsfield units. The corresponding elemental composition of each tissue is then determined on the basis of the International Commission on Radiological Protection Report 23 [35]. The dose grid of AXB was 2.5 mm x 2.5 mm x 2.0 mm in resolution, same as AAA. For each patient, we also exported the planning CT volume. The planning CT volumes were of varying voxel resolutions and volumetric dimensions for different patient cases. We exported the CT and dose files as DICOM-RT files from Eclipse, registered them with DICOM coordinates, and converted them into numeric arrays for the training, validation and testing purposes. Prior to feeding them into the neural network, we rescaled and interpolated both the AAA and AXB doses, as well as the patient-specific CT volumes to a uniform resolution of 1.37 mm x 1.37 mm x 2 mm.

## II.2. Network structure selection

For efficient and accurate dose boosting, we employed a Hierarchically Dense U-Net (HD U-Net) structure [30]. HD U-Net is a combined version of U-Net and DenseNet [36]. Compared with U-Net, HD U-Net uses densely connected layers within each hierarchical level of U-Net, which helps with feature propagation and reuse, and reduces the vanishing gradient issue. Compared with DenseNet, HD U-Net preserves the pooling and up-sampling procedures of U-Net, which are able to capture the global features from the input. Once the HD U-Net structure is set, it trains the same way as U-Net. Quantitative comparisons between the three type of networks have been reported and well-documented in a previous publication [30], showing the advantage of the HD U-Net. As reported, HD U-Net was able to achieve high accuracy with much fewer parameters in the network than U-Net, which reduced the chance of over-fitting. In contrast, DenseNet provided the overall worst results due to its lack of ability in capturing global features. For the supervised training, input channels for the HD U-Net include the Eclipse-calculated AAA dose distribution and the CT volume, and the 'ground-truth' output is the Eclipse-calculated AXB dose distribution. For testing, the output will be the boosted dose. We used patch-based training [37] to balance the size of the training data and the computational resources. We separated the full dose volume (512 x 512 x 128) into patches (patch size: 512 x 512 x 16), and feed each of them individually into the network for training/testing. We then merged the output dose maps into a single volume as the final output. The overall training and testing framework is illustrated in Fig. 1.

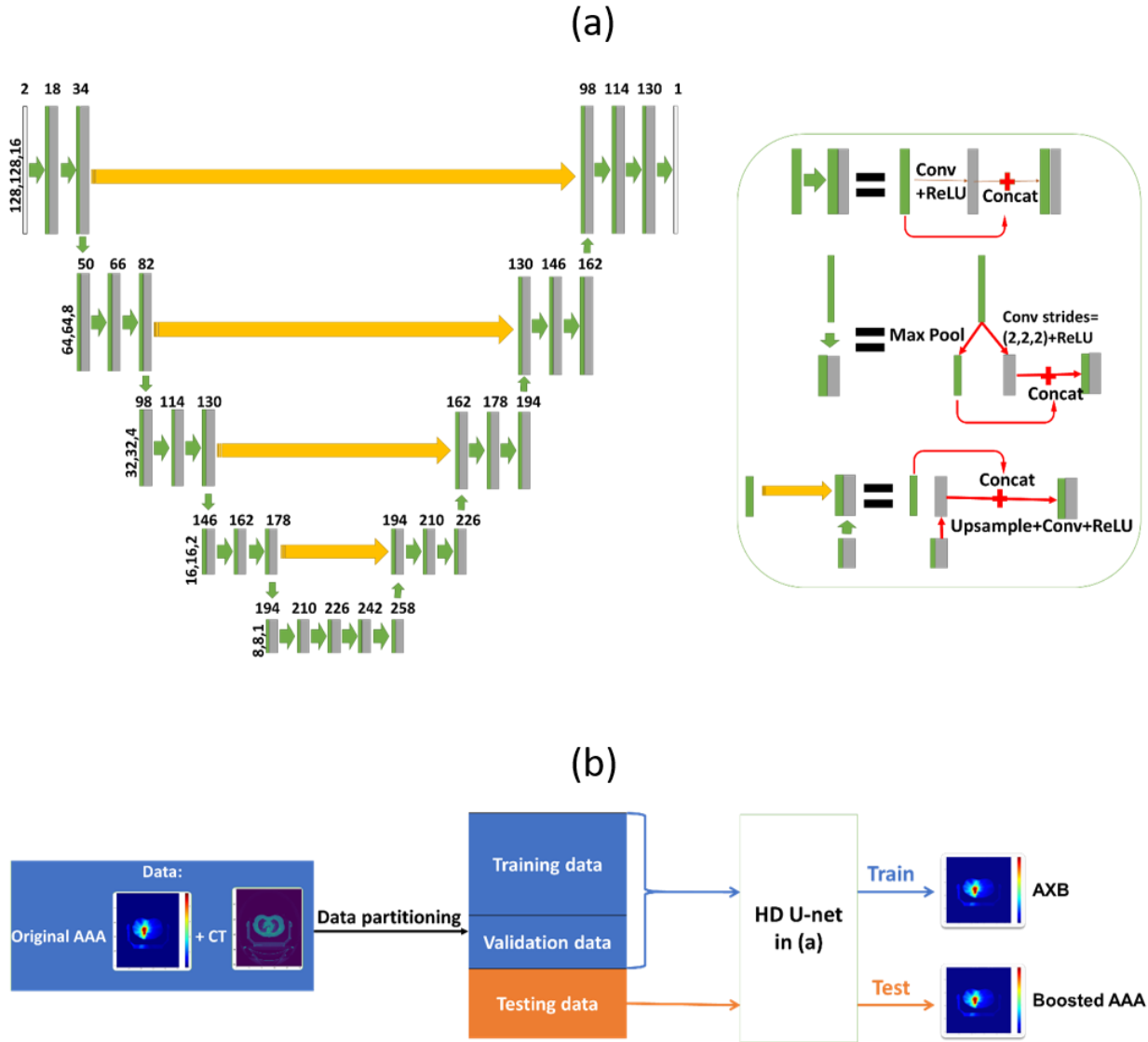

**Figure 1.** (a) General framework of the HD U-Net model for the proposed dose accuracy boosting technique. (b) General training and testing processes where the patient-specific CT and low-accuracy AAA doses serve as the input into the HD U-Net structure, and the high-accuracy AXB doses serve as the 'ground-truth' output for supervised training/validation. Using the trained framework, a new patient-specific CT and low-accuracy AAA dose can be input to obtain a high-accuracy, boosted AAA dose as the output, with its accuracy matching the AXB dose level.

## II.3. Training and testing

Out of the 120 paired, patient-specific AAA-AXB dose maps, we randomly selected 72 sets for training, 18 for validation during training and another 30 for testing. The deep learning model was trained with its hyper-parameters tuned using the validation data set. The HD U-Net contained a hierarchy of five levels to reduce the feature size down to 8 x 8 x 1 at the bottom layer with 2 x 2 x 2 inter-layer max pooling, to learn both local and global features. Within each layer, the convolutional kernel of size 3 x 3 x 3 was implemented with zero padding to maintain the feature size. On the first half of the U-Net, 16 feature maps (filters) were generated in each convolution step. On the remaining half, the number of feature maps of convolution in each layer, except for the very last convolution step, increased by 16 features from the bottom to the top. The last convolution step generated one channel as the final output. Batch normalization was applied after convolution with rectified linear unit (ReLU) operations. The learning rate was set at $10^{-4}$ and the Adam algorithm [38] was selected as the optimizer to minimize the loss defined using the mean-squared-error (MSE). 200 epochs were used for the training in our study, with 100 iterations per epoch. The deep learning model was trained on one NVIDIA Tesla V100 GPU card with 32GB dedicated memory.

To assess the accuracy of the boosted AAA dose map, we visually evaluated its difference with the AXB dose map directly calculated from Eclipse. MSEs were computed between the boosted AAA and AXB dose maps to evaluate their differences. 3D gamma analysis [39] based on both 1%/1mm and 2%/2mm criteria was also performed to quantitatively assess the match between the boosted AAA and AXB dose distributions. We also compared the dose volume histograms (DVHs) of the planning target volume (PTV) and lungs between boosted AAA and AXB doses to evaluate the accuracy of dose conversion. Quantitative dosimetric endpoints, including the $D_{95}$ and $V_{100}$ of the PTV, and $V_{20Gy}$ and $D_{mean}$ of the lungs, were also assessed. The corresponding results between original AAA doses (prior to boosting) and AXB doses were also computed for comparison.

In addition to the proposed network with both original AAA dose and CT as input, we also evaluated a second network using only the original AAA dose as input. The second network was evaluated to assess the potential of directly learning intensity, texture and structural information from the low-accuracy dose maps, to correlate with high-accuracy dose maps for dose boosting. The second network was trained and tested with the same AAA-AXB dose map pairs.

## III. Results

Fig. 2 shows the relative differences between the original AAA and AXB doses (Fig. 2 (b) and Fig. 2 (e)), and between the boosted AAA and AXB doses (Fig. 2 (c) and Fig. 2 (f)), on three views (coronal, sagittal, and axial). Figs. 2 (a) - 2 (c) present a representative non-coplanar static beam plan case, and Figs. 2 (d) – 2 (f) present a volumetric modulated arc plan case. For both cases, the original AAA dose maps show prominent deviations from the reference AXB doses. These dose deviations were substantially reduced in the boosted AAA dose maps.

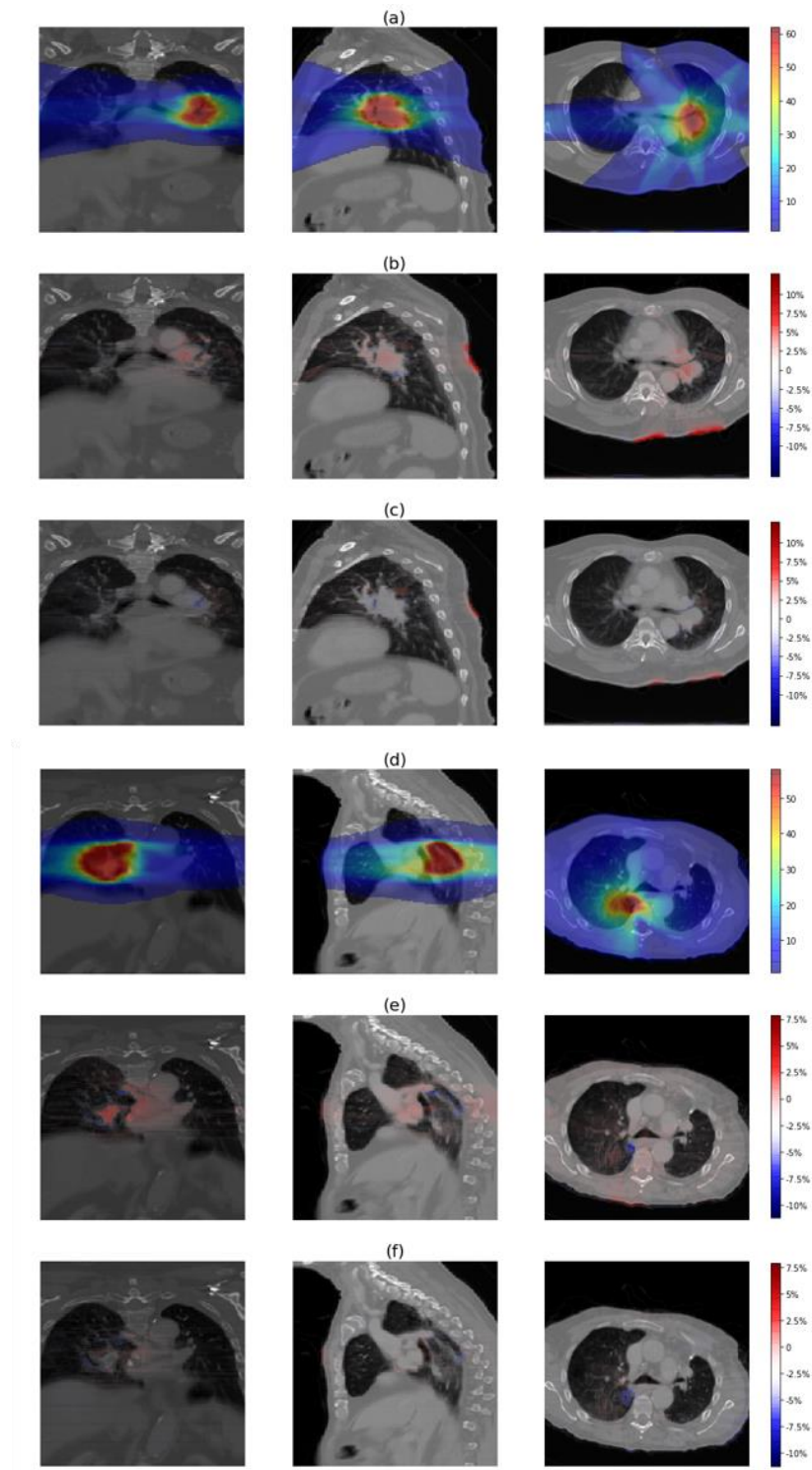

**Figure 2.** (a) The 'ground-truth' AXB dose maps and relative differences between (b) the original AAA and AXB dose maps; (c) the boosted AAA and AXB dose maps, for a 3D non-coplanar static beam plan. (d) The 'ground-truth' AXB dose maps and relative differences between (e) the original AAA and AXB

dose maps; (f) the boosted AAA and AXB dose maps, for a volumetric-modulated arc plan. The 'ground-truth' AXB doses were shown in absolute quantities (Gy). The dose differences were normalized to the plan prescription dose (%).

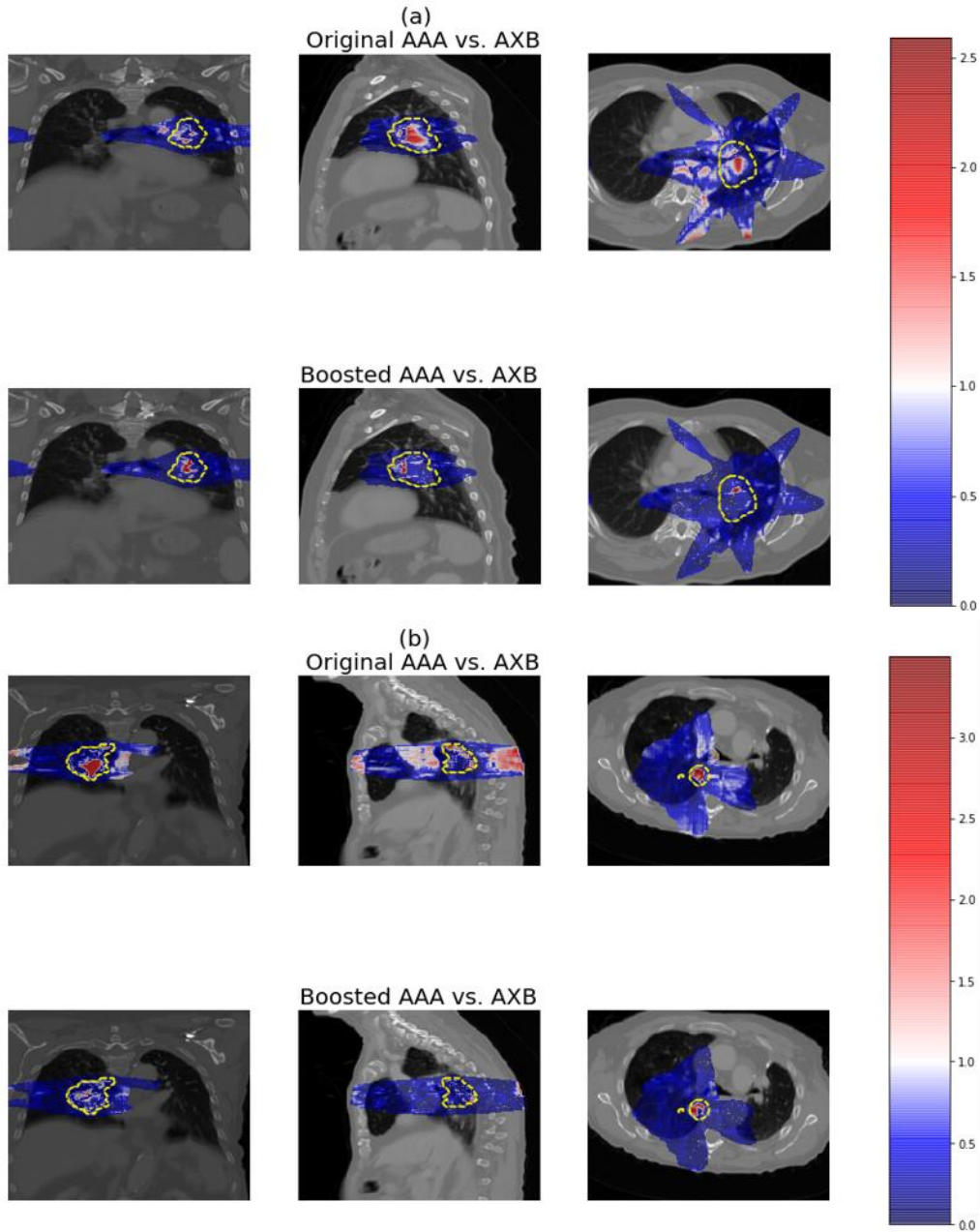

**Figure 3.** Gamma index (1%/1mm) maps between the original AAA and AXB dose distributions, and between the boosted AAA and AXB dose distributions for (a) a 3D non-coplanar static beam plan, and (b)

a volumetric-modulated arc plan. The color bar on the right shows the scale of gamma index and the dashed lines indicate the PTVs. A gamma index > 1 indicates failed gamma test.

In Fig. 3 we showed the gamma index maps between the original AAA and AXB dose distributions, and between the boosted AAA and AXB dose distributions for a 3D non-coplanar static beam plan (Fig. 3 (a)) and a volumetric-modulated arc plan (Fig. 3 (b)), respectively. A stringent criterion (1%/1mm) was used to fully demonstrate the differences between dose maps. The red regions in the map indicated failed gamma index (>1). Large dose discrepancies can be observed on the original AAA gamma index maps, especially around the tumor region. As a comparison, the gamma index maps of the boosted AAA dose have these discrepancies largely removed.

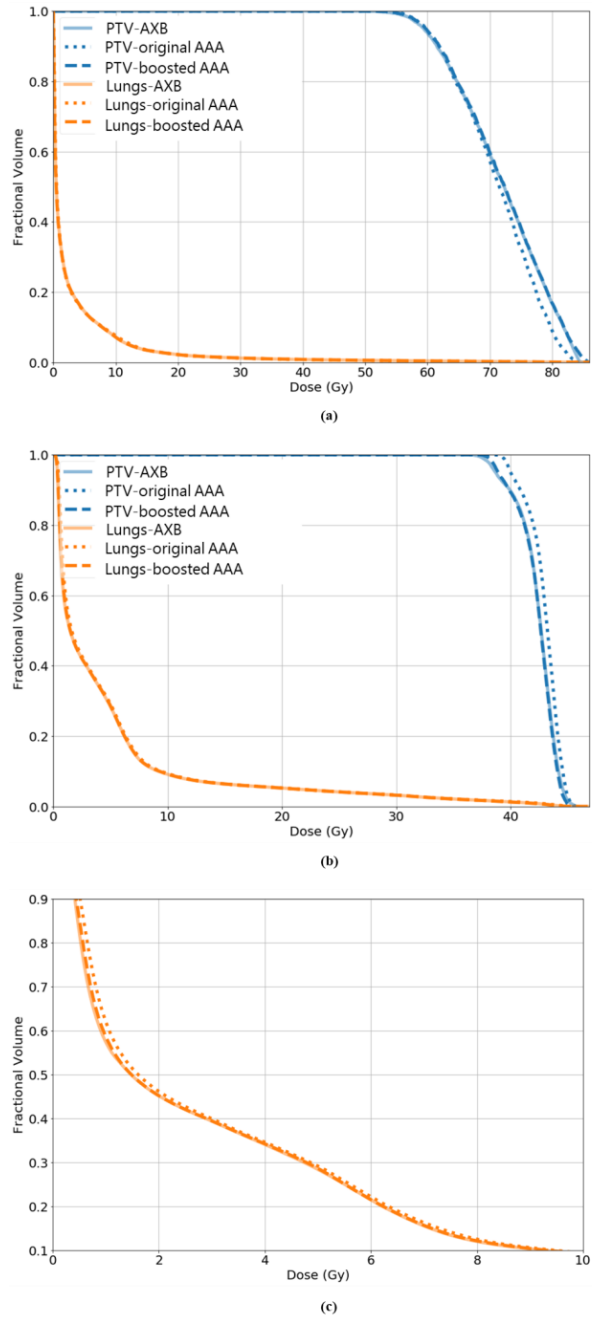

**Figure 4.** The DVH curves of PTV and lungs for (a) a 3D non-coplanar static beam plan, and (b) a volumetric-modulated arc plan. (c) shows the zoomed-in lung DVH curves for the volumetric-modulated arc plan. The solid, dashed and dotted lines correspond with the AXB, boosted AAA, and original AAA dose maps, respectively.

Fig. 4 compared the DVH curves between the original AAA, AXB, and boosted AAA dose distributions, for both PTV and lungs. The DVH curves of boosted AAA doses matched well with those of AXB doses, while substantial discrepancy could be observed between DVH curves of the original AAA and AXB doses.

**Table 1.** Quantitative comparisons between the original AAA and AXB doses, and between the boosted AAA (w/ and w/o CT as input) and AXB doses. MSE: mean-squared-error. RX: prescription. The results of the 30 testing patient cases were included in the analysis.

| Dose maps | Gamma passing rates (%) | | MSE of voxels with dose > 5% RX dose | % of voxels with dose deviations > 3% RX dose |
|---|---|---|---|---|
| | 2mm/2% | 1mm/1% | | |
| Original AAA vs. AXB | 98.4 ± 1.5 | 87.8 ± 9.0 | 0.31±0.21 | 2.01 ± 1.19 |
| Boosted AAA w/o CT vs. AXB | 99.3 ± 0.7 | 94.6 ± 2.8 | 0.15±0.10 | 0.85 ± 0.53 |
| Boosted AAA vs. AXB | 99.8 ± 0.4 | 97.6 ± 2.4 | 0.11±0.05 | 0.46 ± 0.46 |

In Table 1, the boosted AAA doses demonstrated substantially improved match to the AXB doses, with average (±s.d.) gamma passing rate (1 mm/1%) 97.6% ±2.4%, compared to 87.8% ±9.0% for the original AAA doses. Using a less strict criterion (2 mm/2%) yielded 99.8% ±0.4% for the boosted AAA doses, compared to 98.4% ±1.5% for the original AAA doses. The corresponding average MSE was 0.11 ±0.05 between the boosted AAA and AXB doses, compared to 0.31 ±0.21 between the original AAA and AXB doses. The boosted AAA doses (w/o CT as input) were of accuracy in between the original AAA doses and the boosted AAA doses. Note that since this paper focuses on developing a network using CT as one of the input, if not specifically mentioned, boosted AAA doses refer to those obtained from this network.

**Table 2.** Comparisons between the original AAA, AXB, and boosted AAA doses in terms of $D_{95}$ of PTV, $V_{100}$ of PTV, $D_{mean}$ of lungs, and $V_{20Gy}$ of lungs. The results of the 30 testing patient cases were included in the analysis.

| | PTV | | | | | |
|---|---|---|---|---|---|---|
| | $D_{95}$ (Gy) | | | $V_{100}$ | | |
| | Original AAA | AXB | Boosted AAA | Original AAA | AXB | Boosted AAA |
| Mean | 45.00 | 44.43 | 44.40 | 94.48% | 89.40% | 89.71% |
| s.d. | 7.82 | 7.94 | 7.84 | 0.98% | 10.73% | 8.50% |
| | Lungs | | | | | |
| | $D_{mean}$ (Gy) | | | $V_{20Gy}$ | | |
| | Original AAA | AXB | Boosted AAA | Original AAA | AXB | Boosted AAA |
| Mean | 4.49 | 4.42 | 4.41 | 5.89% | 5.84% | 5.85% |
| s.d. | 2.61 | 2.57 | 2.58 | 4.43% | 4.37% | 4.39% |

In Table 2, dosimetric comparison results were reported in terms of $D_{95}$ of PTV, $V_{100}$ of PTV, $D_{mean}$ of lungs, and $V_{20Gy}$ of lungs. Wilcoxon signed-rank tests with the Bonferroni corrections revealed that the AXB was significantly distinct from the original AAA in $D_{mean}$ of lungs (p-value $< 10^{-5}$), $D_{95}$ of PTV (p-value = 0.001), and $V_{100}$ of PTV (p-value = 0.002). In contrast, the boosted AAA did not statistically differ from the AXB (p-value = 0.094 for $D_{mean}$ of lungs, p-value = 0.642 for $D_{95}$ of PTV, and p-value = 0.417 for $V_{100}$ of PTV). No significant differences were found for $V_{20Gy}$ of lungs, either for the boosted AAA and AXB pair, or the original AAA and AXB pair. Therefore, the boosted AAA matched better with AXB than the original AAA doses, which provided more accurate target coverage and OAR sparing information.

## IV. Discussion

Accurate dose calculation and reporting are key to effective and safe radiation therapy. A dose calculation algorithm with both high accuracy and high efficiency is much desired in today's clinic for treatment planning and dose reporting. Our study demonstrated that through a deep learning framework, we could achieve a high-accuracy dose map by boosting from a low-accuracy dose map with patient-specific anatomical CT information, to successfully overcome the trade-off between computational speed and accuracy. The training of our deep learning model takes around ~48 hours for 200 epochs. Currently, the inference of our deep learning model takes on average 19 seconds, with a standard deviation of 3 seconds. The relatively long inference time is mostly due to the memory limit of our GPU hardware. Under the memory limit, we have to use a patch-based strategy to boost the AAA dose patch by patch, and then merge the results together for a full dose volume. With a larger GPU memory, we can boost the whole AAA dose volume by running the model only once, which takes roughly 1-2 seconds. In addition, in this study we performed dose boosting on an AAA dose volume of 512 x 512 x 128 voxels, while the meaningful dose cloud only occupies a much smaller region within. We may trim the AAA dose volume to remove the irrelevant regions of minimal doses, and apply the deep learning model to the remaining dose cloud only. By this way, we can also remove the need of patch-based model inference and accelerate dose boosting to ~1 second. To put the dose boosting time into context, for our AAA dose and AXB dose calculations, we found the AXB calculation was around 1.5 to 5 times slower than the AAA calculation, depending on the treatment technique and beam arrangement of each plan. In terms of seconds/minutes, AXB can be ~30 seconds to more than two minutes slower than AAA. In general, AXB tends to be much less efficient for plans with static-gantry beams, which are frequently used in our clinic for lung treatments. Our dose boosting scheme, which can potentially be executed within 1 second, will significantly improve the dose calculation efficiency.

In this study, we trained, validated and tested a voxel-wise dose boosting model on 120 in-house lung cancer patient cases using a Hierarchically Dense U-Net. Visual comparisons of dose differences showed major improvements in the boosted AAA doses, for areas both within the tissues and along tissue interfaces (Fig. 2). In comparison, the original AAA lacks accuracy in calculating doses at multiple regions, where electron densities are changing rapidly and invalidate the kernel scaling approach it applied to account for tissue inhomogeneity. Gamma index distribution maps shown in Fig. 3 also confirmed the improvement of accuracy in boosted AAA doses. The structural-specific DVH curves and dosimetric endpoints also demonstrated that the boosted AAA doses matched well with the AXB doses (Fig. 4, Table 2), and provided more accurate target coverage and OAR sparing information. In general, the conversion model yielded ~98% gamma passing rate between the boosted AAA and AXB doses for 1%/1mm gamma analysis and ~100% gamma passing rate for 2%/2mm gamma analysis, showing almost perfect match. In comparison, the corresponding results were only ~88% and ~98% for the original AAA doses (Table 1).

In our developed network, we used both CT images and a low-accuracy dose map as input to derive a high-accuracy dose map. We also evaluated the feasibility of directly using the low-accuracy dose map (w/o CT) to correlate with the high-accuracy dose map for dose boosting (Table 1). It can be observed that the network without using CT as input also helps to boost the AAA dose to match better with the AXB dose than the original AAA dose. However, it is also evident that the boosted AAA dose with CT as input has its accuracy best matched with the AXB dose. With the CT images providing electron density information, the HD U-Net will be better informed of potential inaccuracies in the original AAA dose maps through interpreting the CT density information, to further improve the accuracy of dose boosting.

In current clinical practice, convolution/superposition algorithms like AAA remain the main dose calculation engine for clinical TPS. To improve the dose calculation accuracy, AXB has been introduced into clinics as an alternative to full Monte Carlo simulation, but its clinical use is still limited. Potential hurdles include the acquisition cost, the lack of resources, the lack of experience to commission a new AXB engine, and often, the physicians' familiarity and reliance with AAA doses upon which the clinical experience was accumulated. However, it is always highly desirable to use the most accurate dose distribution to correlate with clinical outcomes, in order to derive a reliable and reproducible dose-outcome relationship to benefit and guide future practices. The dose boosting model developed by us can contribute to this goal, as it could be easily adopted by clinics to prospectively/retrospectively convert the original AAA doses to AXB-quality doses. The potential fast-speed conversion (within seconds) introduces minimal interference towards current clinical workflow. The model could be incorporated into the TPS's application programming interface (API) (e.g. ESAPI) [40], to enable direct dose conversion with a one-click solution to allow physicians to evaluate and compare the original AAA and boosted AAA doses side-by-side. Currently, our model is built and run on Python, an interpreted, general-purpose programming language. To run our model in Eclipse, one way is to compile our Python code into an executable, and call the executable from within Eclipse using scripts. There are also ongoing developments on a Python interface for ESAPI, which could make an alternative path to run our model within Eclipse. The model could also be run in batch in the background to accumulate boosted AAA data for further dose-clinical outcome analysis. The dose conversion model also does not incur costs associated with proprietary dose calculation algorithms, which can potentially benefit less resourceful, under-served cancer centers. We believe the accuracy and convenience offered by the dose map conversion framework will help physicians to make a more informed decision when evaluating treatment plans, and ultimately benefit the current radiotherapy practice with more accurate dose calculations especially in heterogeneous regions [41].

In this study, all the evaluated patient data sets are acquired from our institution. The developed model needs to be further evaluated on patient data sets from other institutions to assess its transferability. Though no substantial model adjustments are expected, some slight model tweaking might be needed to accommodate the inter-institutional variations. Transfer learning, an artificial intelligence technique that allows easy model adaptation based on limited new data samples, may help in the case [28]. Similarly, all our data are currently trained on lung cancer patients, other clinical scenarios where the inhomogeneity might affect the dose accuracy, such as larynx and pelvis [42], also warrant additional model evaluation and potential model fine-tuning.

In addition to our study, there are other groups working on projects to uncover the potential of deep learning in dose accuracy enhancement. A recent work by [1] uses parallel U-Net branches to boost Monte-Carlo dose calculation accuracy. However, this method is currently limited to boosting doses calculated by a single beamlet, as compared to the full 3D dose map boost achieved by our method. In addition, the network developed by [1] works by boosting low-accuracy Monte-Carlo doses (fewer events: faster calculation, more noise, more uncertainty) to high-accuracy Monte-Carlo doses (more events: slower calculation, less noise, less uncertainty), essentially an intra-algorithm conversion (between same Monte-Carlo type algorithms). In comparison, our method can boost the doses from one type of algorithm to another (AAA to AXB in our study), allowing inter-algorithm conversion. Another study by [2] is also converting between low-accuracy and high-accuracy Monte-Carlo doses for de-noising, which may not be readily applicable to non-Monte-Carlo based dose calculation algorithms which are dominant in current clinical TPSs.

In summary, it is shown in this study that with the power of deep learning, we can uncover a mapping scheme between low-accuracy and high-accuracy dose maps, using patient anatomical structure maps and intensity distributions as guidance. The "low" and "high" are defined only relatively in the context of the two algorithms under study, and should not be interpreted in an absolute fashion. We tested this hypothesis through developing and evaluating a dose boosting framework between AAA and AXB dose maps. This framework can be readily extended to other potential pairs of dose maps, the relative accuracy difference of which may be more pronounced. The relative accuracy levels of different algorithms can be determined through the four types of algorithms defined in the AAPM Task Report No. 85 [1], or the 'a' to 'c' categories stratified by [45]. It will be of interest to test the framework in boosting pencil-beam convolution doses to Monte-Carlo doses, which might have a positive impact on treatment plan optimization as less-accurate pencil-beam convolution-type calculations are usually used within optimization to promote efficiency. It also remains to be investigated how robust the current framework will be to boost a rudimentary algorithm (such as the correction-based dose calculation algorithm) to a high-accuracy algorithm like Monte Carlo. Additional information, like the treatment plan itself, could be potentially fed into the AI framework to improve the dose boosting accuracy if needed.

## V. Conclusions

A deep learning-based framework was developed to convert dose maps to boost their accuracy towards the AXB level in a time frame of seconds. The developed method allows dose computation with both high efficiency and high accuracy, potentially benefiting other applications including retrospective dose analysis (especially for heterogeneous regions), fast plan optimization, and secondary dose calculations and plan checks. Future studies involving other dose algorithms, anatomical sites and treatment centers are warranted to further evaluate the efficacy and robustness of the developed framework.

## VI. Conflict of Interest Statement

No conflict of interest.

## VII. Acknowledgements

The authors would like to thank the National Institutes of Health (NIH), the Cancer Prevention and Research Institute of Texas (CPRIT) for providing support through grants 1R01CA237269 (NIH), IIRACA RP160190 (CPRIT), and IIRA RP150485 (CPRIT). The authors would like to thank the Varian Medical Systems for providing research grant support. The authors would also like to thank the UT Southwestern Medical Center for providing research support through Radiation Oncology departmental seed grants.

The authors would like to thank Qiming Yang for providing the python-based DICOM structure extraction tool.